\pdfoutput=1
\documentclass[letterpaper,twocolumn,showpacs,
 preprintnumbers,amsmath,amssymb,floatfix,nofootinbib]{revtex4}

\usepackage{graphicx}
\usepackage{bm}
\usepackage{url,multirow,xcolor}

\begin{document}

\title{Testing Skyrme energy-density functionals with the QRPA in low-lying
vibrational states of rare-earth nuclei}

\author{J.\ Terasaki}\altaffiliation[Present address: ]{Center for Computational Sciences, University of Tsukuba, Tsukuba, 305-8577, Japan}
\affiliation{Department of Physics and Astronomy, University of North Carolina,
Chapel Hill, NC 27599-3255}
\author{J.\ Engel}
\affiliation{Department of Physics and Astronomy, University of North Carolina,
Chapel Hill, NC 27599-3255}

\begin{abstract} 
Although nuclear energy density functionals are determined primarily by fitting
to ground state properties, they are often applied in nuclear astrophysics to
excited states, usually through the quasiparticle random phase approximation
(QRPA).  Here we test the Skyrme functionals SkM$^\ast$ and SLy4 along with the
self-consistent QRPA by calculating properties of low-lying vibrational states
in a large number of well-deformed even-even rare-earth nuclei.  We reproduce
trends in energies and transition probabilities associated with
$\gamma$-vibrational states, but our results are not perfect and indicate the
presences of multi-particle-hole correlations that are not included in the
QRPA.  The Skyrme functional SkM$^\ast$ performs noticeably better than SLy4.
In a few nuclei, changes in the treatment of the pairing energy functional have
a significant effect.  The QRPA is less successful with ``$\beta$-vibrational''
states than with the $\gamma$-vibrational states.  
\end{abstract}

\maketitle
\section{Introduction}
Modern supercomputers are making the quantitative theoretical treatment of
nuclear structure increasingly common.  In light nuclei, Greens-function Monte
Carlo methods \cite{Pie04,Pie01} and the no-core shell model
\cite{Neg10,Var09} yield accurate ab initio results, and in medium-mass nuclei
the coupled cluster method \cite{Hag07,Dea04} is proving successful.  In nuclei
with $A>50$, techniques related to density-functional theory (DFT) \cite{Dob10}
are the state of the art.  Accuracy, at least for ground-state properties, is
limited only by the quality of the functionals, which are continually improving
\cite{Kor10}.  

One advantage of DFT is its applicability to nearly all heavy nuclei.  Such
flexibility is particularly important for nuclear astrophysics, which attempts
to explain the synthesis of all the elements.  Another advantage is a natural
extension, through the self-consistent quasiparticle random phase approximation
(QRPA) to excitations.  Excited states are as important as ground states in
many nucleosynthetic reactions, and so Goriely et al.~\cite{Gor02}, for
example, used the QRPA to compute radiative neutron capture in a wide range of
nuclei.  Such calculations, however, have generally ignored deformation, or
treated it in a crude way. The logical next step is to take the effects
of deformation into account in a self-consistent fashion. 

Fortunately, self-consistent QRPA calculations in heavy deformed nuclei
are now becoming possible.  Recently, we developed a scaled parallel
Skyrme-QRPA code \cite{Ter10} for arbitrary axially-deformed (parity
conserving) even-even nuclei.  Our code is one of the few \cite{Per11} to treat
heavy deformed nuclei in the QRPA without simplification. (For other
calculations, including those in lighter nuclei and those in the RPA, the
spherical QRPA, and separable approximations, see the work cited in
Ref.~\cite{Ter10} and, e.g., the more recent Ref.\ \cite{Yos11}.)  In this
paper, we present calculations with two Skyrme energy-density functionals of
properties of low-energy vibrational states in rare-earth nuclei.  As promised
in Ref.\ \cite{Ter10}, we discuss the performance of both the functionals and
the QRPA.

In Sec.~\ref{sec:test} below we list the nuclei that we explore and present
technical information about our calculations.  In Sec.~\ref{sec:gamma} we show
results for $\gamma$-vibrational states and discuss the performance of the
Skyrme QRPA, which we compare with methods used in earlier calculations.
Sec~\ref{sec:beta} treats ``$\beta$-vibrational'' states\footnote{We put the
term in quotes to indicate that many of those states are not purely vibrational
\cite{Gar01}.} briefly, and Sec.~\ref{sec:conclusion} is a conclusion.  An
appendix presents equations for two-body matrix elements of the Coulomb-direct
interaction and discusses computational efficiency. 

\section{\label{sec:test} Selection of Nuclei and Method of Calculation}

\begin{figure}
\includegraphics[width=7cm]{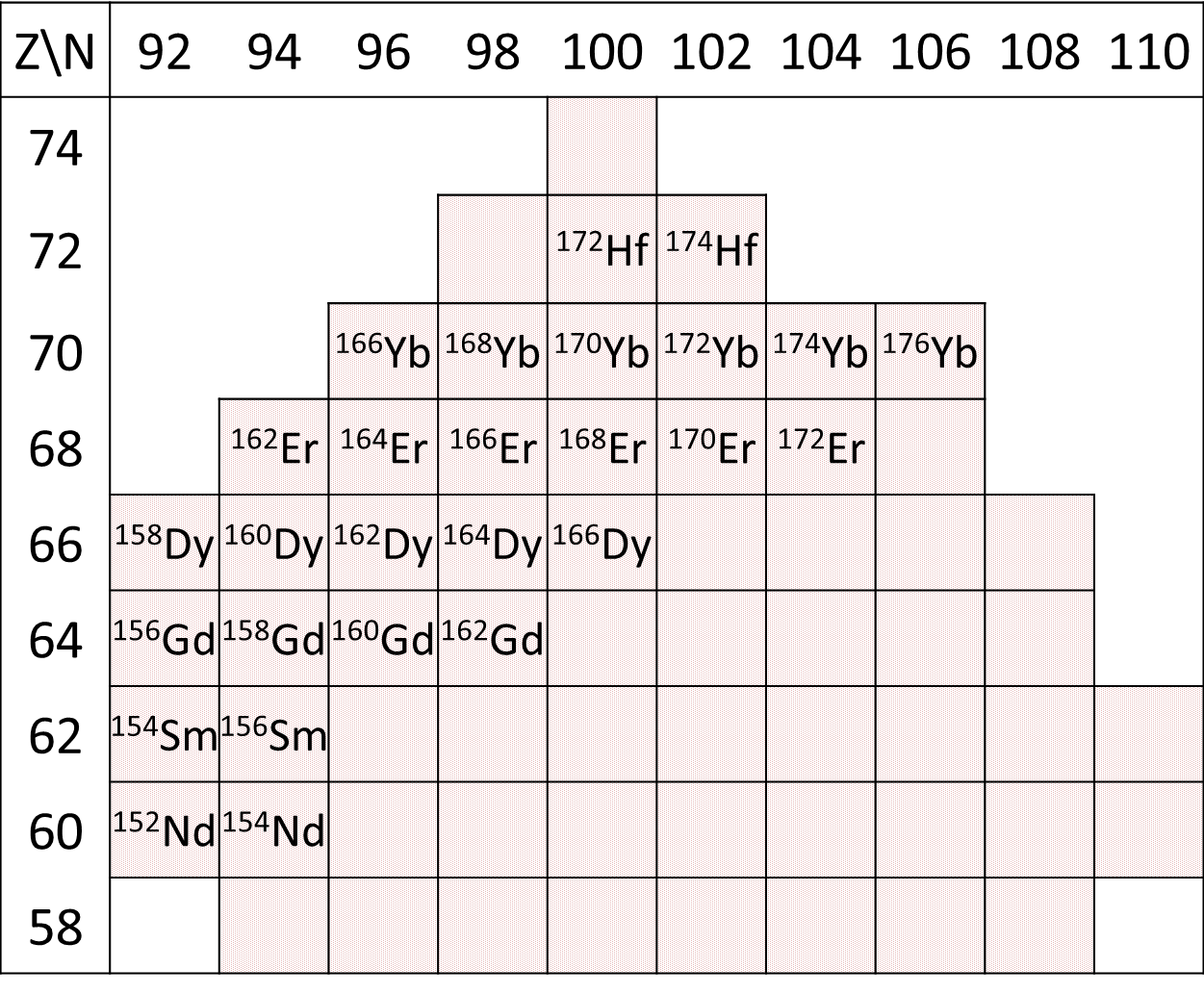}%
\caption{\label{fig_nuclei_calculated} (Color online) Rare-earth region of the
isotopic chart. Shaded area shows nuclei with deformation $\beta\ge 0.3$ in HFB
calculations with SkM$^{\ast}$ (unpublished, see Ref.~\cite{Sto03} for SLy4
which gives a similar result).  We calculate energies and transition probabilities 
of $\gamma$-vibrational states for all nuclei whose isotopic symbols appear in
the figure.  Squares without symbols correspond to nuclei for which
experimental data on $\gamma$ vibrations are not in Ref.\ \cite{nndc}.}
\end{figure}

The vibrational states we examine are all in rare-earth nuclei. The advantage
of this region of the isotopic chart is the abundance of reliable experimental
data \cite{Osh95, Sug93, Fah92, Fah92b, Kot90, Bur88, Ich87, Wal83, Ron82,
Cre81, McG81, Rie79, McG78, Wol77, Ron77, Rei74, Bak74, Oeh74, Car73, Bem73,
Dom72, Gil71, Cha71, Eji71, Gro68, Vej68, Blo67, Yos65, Nat60}, accumulated
over the last half century.  Multiple results exist for many of the nuclei and
there are few serious discrepancies.  In addition, the 
large deformation of
many of the rare earths make them better candidates for a successful QRPA
treatment than transitional nuclei, which tend to be soft.  We choose the 27
nuclei shown in Fig.~\ref{fig_nuclei_calculated} for our calculation.  They are
all axially symmetric and well deformed, with $\beta\geq$ 0.3, and for all but
a few the energies of their $\gamma$-vibrations ($K^\pi=2^+$, the second or
third $J^\pi=2^+$ states) have been measured and appear in Ref.~\cite{nndc}.
We calculate $\gamma$-vibrational energies and E2 excitation strengths in all
27 nuclei with the Skyrme functionals SkM$^\ast$ \cite{Bar82} and SLy4
\cite{Cha98}, and in a few nuclei we do the same for ``$\beta$-vibrational"
states ($K^\pi=0^+$) with SkM$^\ast$.  We use the traditional volume-pairing
energy functional \cite{Ter05} for simplicity.

Our procedure has two steps: a Hartree-Fock-Bogoliubov (HFB) calculation with
the Vanderbilt HFB code \cite{Bla05}, and a QRPA calculation that uses the
results of the HFB run.  Both steps use B splines \cite{Deb78, Nur89, Sch07} to
represent wave functions on a 42 by 42 cylindrical mesh
with $ 0 \leq z, \rho \leq 20$ \textrm{fm}.  We use box boundary conditions to
discretize the continuum, and introduce a quasiparticle cutoff energy
$E_\textrm{cut}$ of 60 MeV or 200 MeV in the HFB calculation to limit the set
of quasiparticle wave functions that determine the density and the pairing
tensor.  The two cutoffs require different pairing strengths, which we adjust
via the three-point formula \cite{Boh69} so as to reproduce the pairing gaps of
$^{172}$Yb (obtained from experimental masses).  In the other nuclei, this
procedure usually reproduces overall pairing gaps to within $\pm$150 keV. We
restrict the $z$-component of the angular momentum of the wave functions to be
less than or equal to 19/2 $\hbar$.

Next we transform the quasiparticle wave functions to the canonical-basis and
introduce two cutoff occupation probabilities
$(v^\textrm{cut}_\textrm{pair})^2$ and $(v^\textrm{cut}_\textrm{ph})^2$,
used also in our prior work \cite{Ter05,Ter08,Ter10}, to truncate the two-canonical-quasiparticle
basis in which we construct the QRPA Hamiltonian matrix.  We take
$((v^\textrm{cut}_\textrm{pair})^2, ((v^\textrm{cut}_\textrm{ph})^2) =
(10^{-4}, 10^{-8})$ for $E_\textrm{cut}=60$ MeV, and $(10^{-3}, 10^{-6})$ for
$E_\textrm{cut}=200$ MeV in the $\gamma$-vibration calculation with
SkM$^\ast$.  Those values make the dimension of the two-canonical-quasiparticle
basis about 22000 in $^{172}$Yb, a number that is large enough to yield a
convergent result.  In the other rare-earth nuclei, the dimension ranges from
19000 to 28000.

Spurious states associated with particle-number conservation make the necessary
space much larger for ``$\beta$-vibrations." There we use
$((v^\textrm{cut}_\textrm{pair})^2, ((v^\textrm{cut}_\textrm{ph})^2) = (
10^{-4}, 10^{-6} )$ with $E_\textrm{cut}$ = 200 MeV, values with which the
dimension of the two-canonical-quasiparticle space is 60000 to 75000. Even with
this large dimension, however, the spurious state does not separate perfectly
and we present results only for cases in which the separation is good.

Deriving the QRPA equations for an axially-symmetric system is tedious but not
difficult and can be done by starting from the general equations in, e.g.,
Ref.~\cite{Ter05}. In the appendix, therefore, we display only our
representation of the Coulomb-direct matrix elements.  These require more
numerical effort than matrix elements of a $\delta$-interaction, and so benefit more from a
computationally efficient procedure.

\section{\label{sec:gamma} {$\bm{\gamma}$}-vibrations}

\subsection{Energies and transition strengths}

\begin{figure*}[t] 
\includegraphics[width=12cm]{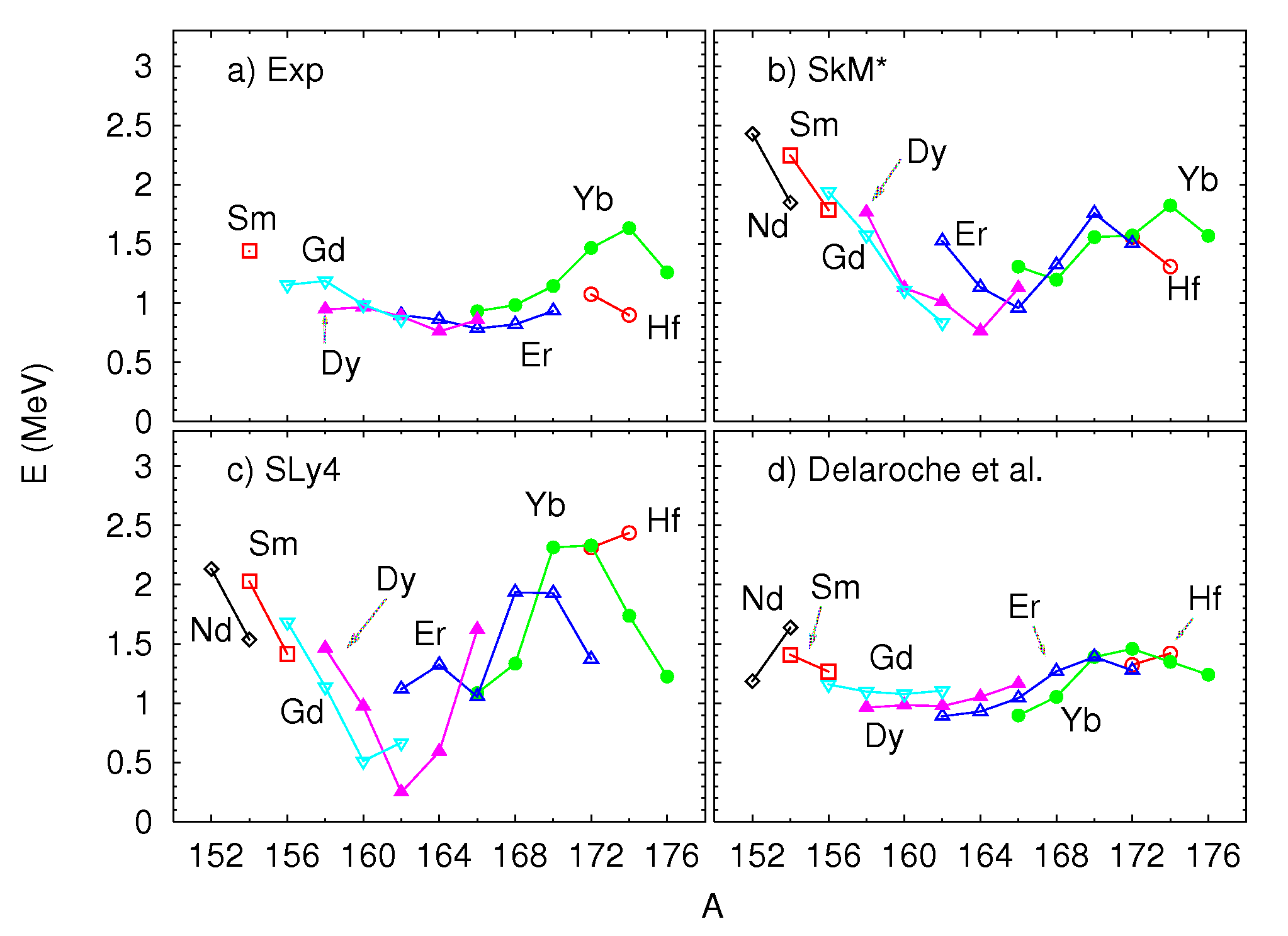}%
\caption{\label{fig_e_exp_skmstar_sly4_delaroche} (Color online) Energies of
$\gamma$-vibrational states from a) experiment \cite{nndc}, b) SkM$^\ast$, c)
SLy4, and d) Delaroche et al.~\cite{Del10}.  }
\end{figure*}
\begin{figure*}
\includegraphics[width=12cm]{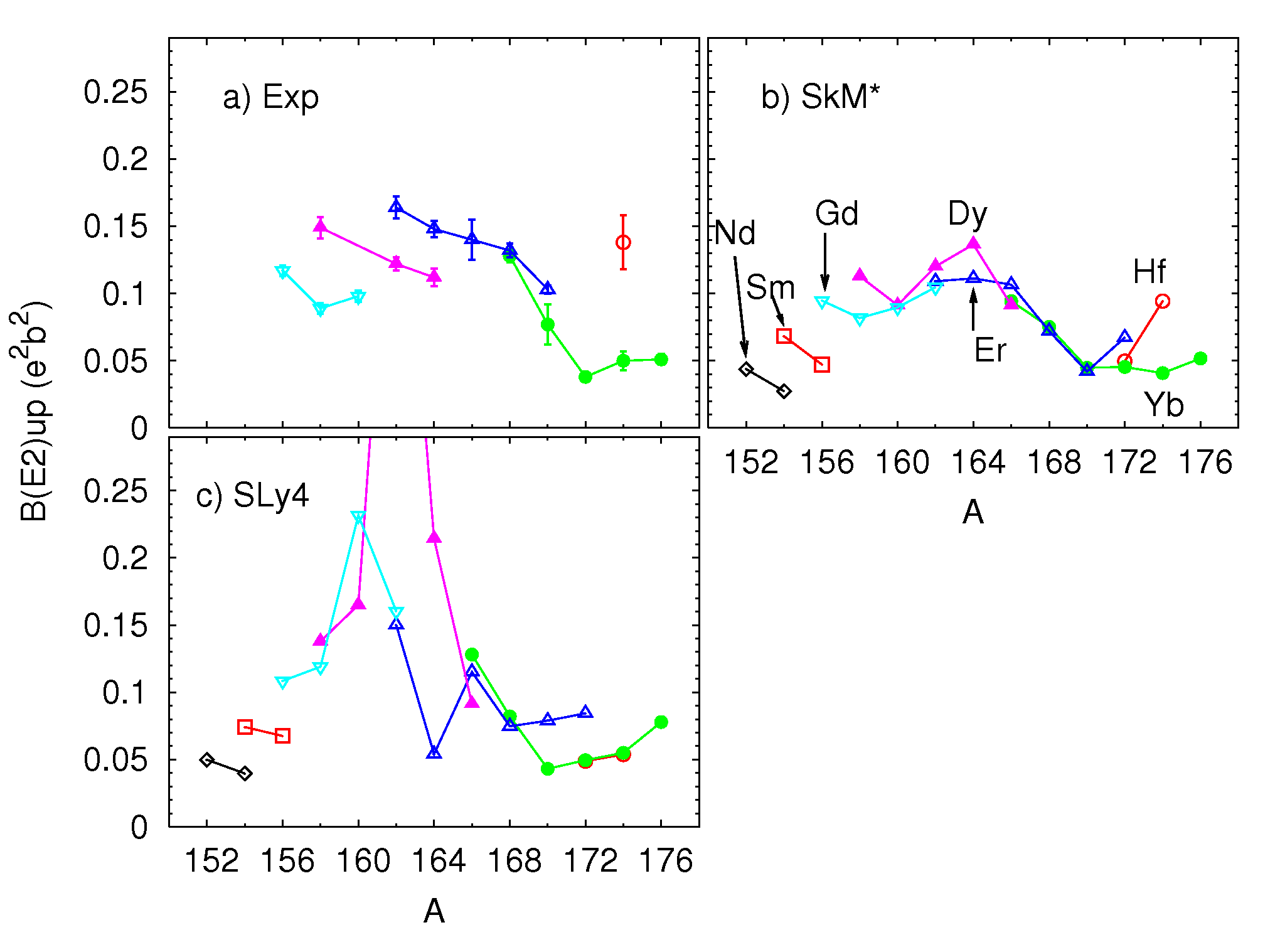}%
\caption{\label{fig_be2_exp_skmstar_sly4_delaroche} (Color online)
$B(E2;0^+_\textrm{gs}\rightarrow 2^+_\gamma)$ corresponding to
Fig.~\ref{fig_e_exp_skmstar_sly4_delaroche}. The value for $^{162}$Dy in c) is
0.562 $e^2\textrm{b}^2$. This figure has no panel d) because the results from
the calculation of Delaroche et al.~\cite{Del10} are not published.  We include
only those experimental data that are labeled $\gamma$-vibrations in Ref.\
\cite{nndc}.  The symbols for particular isotopic chains are the same in each
panel.}
\end{figure*}

Figure \ref{fig_e_exp_skmstar_sly4_delaroche} shows measured $\gamma$-vibration
energies alongside the results of our two QRPA calculations and a
collective-model calculation (with parameters determined from the Gogny energy
functional) by Delaroche et al.\ \cite{Del10}.  In all the plots the minimum
energy occurs around $A$=162.  The minimum in the Dy and Er isotopes is at
$N=98$, both in the data and the SkM$^\ast$.  The Yb isotopes are particularly
well reproduced by the SkM$^\ast$ calculation. But the QRPA calculations show a
stronger $A$-dependence than the data, with the SLy4 results showing the
strongest dependence.  And overall, neither of the QRPA calculations is as good
as that of Ref.\ \cite{Del10}. 

Figure \ref{fig_be2_exp_skmstar_sly4_delaroche} shows $E2$ transition strengths
$B(E2;0^+_\textrm{gs}\rightarrow 2^+_\gamma)$, hereafter denoted
$B(E2)\!\!\uparrow$, for the same isotopes. Overall, the calculations reproduce
the data reasonably well, except in $^{162}$Dy with SLy4, and again are
particularly good in the Yb isotopes.  As before, SkM$^\ast$ is noticeably
better than SLy4.  The energies and $B(E2)\!\!\uparrow$'s in our calculations
are anticorrelated in general, a feature expected of harmonic vibrations.  On
the other hand, the experimental $B(E2)\!\!\uparrow$'s in Er decrease
monotonically with $A$, even though the dependence of the energy is slightly
parabolic.  

To characterize the performance of the two functionals statistically, we
introduce, following Refs.~\cite{Ter08, Ber07} the measures
\begin{equation}
R_E = \ln\left( E_\textrm{cal}/E_\textrm{exp} \right)
\end{equation}
and
\begin{equation}
R_Q = \ln \sqrt{ B(E2)\!\!\uparrow_\textrm{cal}/B(E2)\!\!\uparrow_\textrm{exp}
} \,,
\end{equation}
where $E_\textrm{cal}$ and $E_\textrm{exp}$ are the calculated and experimental
energies of the $\gamma$-vibrational state. The results are in
Tab.~\ref{tab_rerq}.  SLy4 actually does better than SkM$^\ast$ in the
averages, but gives much larger dispersions.

\begin{table}[t]
\caption{\label{tab_rerq}%
Average of $R_E$ ($\bar{R}_E$), dispersion of $R_E$ ($\sigma_E$), and the same
for $R_Q$. }
\begin{ruledtabular}
\begin{tabular}{lcccc}
 & $\bar{R}_E$ & $\sigma_E$ & $\bar{R}_Q$ & $\sigma_Q$ \\
\colrule SkM$^\ast$ & 0.28 & 0.18 & $-$0.13 & 0.14\\
SLy4       & 0.20 & 0.50 & $-$0.004 & 0.31\\
\end{tabular}
\end{ruledtabular}
\end{table}
 
Table \ref{tab_rerq_sph} shows the statistical measures for the spherical
nuclei treated in Ref.\ \cite{Ter08} and for the subset of those nuclei that
exhibit ``low softness." (Some of the other nuclei in Ref.~\cite{Ter08} are
transitional.) There are far more nuclei in the spherical data set than in the
deformed rare-earth set, so it is hard to make a precise comparison of
performance.  But deformation does not appear to affect it significantly.
 
\begin{table}[t]
\caption{\label{tab_rerq_sph}%
Statistical measures for the spherical nuclei and for the subset with low
softness from Ref.~\cite{Ter08}.  $R_Q$ and $\sigma_Q$ were not calculated
separately for the low-softness nuclei in that paper because of a lack of
$E2$ data.}
\begin{ruledtabular}
\begin{tabular}{llcccc}
                &    & $\bar{R}_E$ & $\sigma_E$ & $\bar{R}_Q$ & $\sigma_Q$ \\
\colrule \multirow{2}{*}{All} & SkM$^\ast$ & 0.11 &0.44 & -0.29 & 0.53 \\
                              & SLy4       & 0.33  &0.51 & -0.32 & 0.42 \\
\multirow{2}{*}{Low Softness} & SkM$^\ast$ & 0.27  &0.35 & ---   & ---  \\
                              &SLy4        & 0.47  &0.48 & --- & ---    \\
\end{tabular}
\end{ruledtabular}
\end{table}
 
\subsection{$\bm{N}$- and $\bm{Z}$-dependence}
 
\begin{figure*}
\includegraphics[width=12cm]{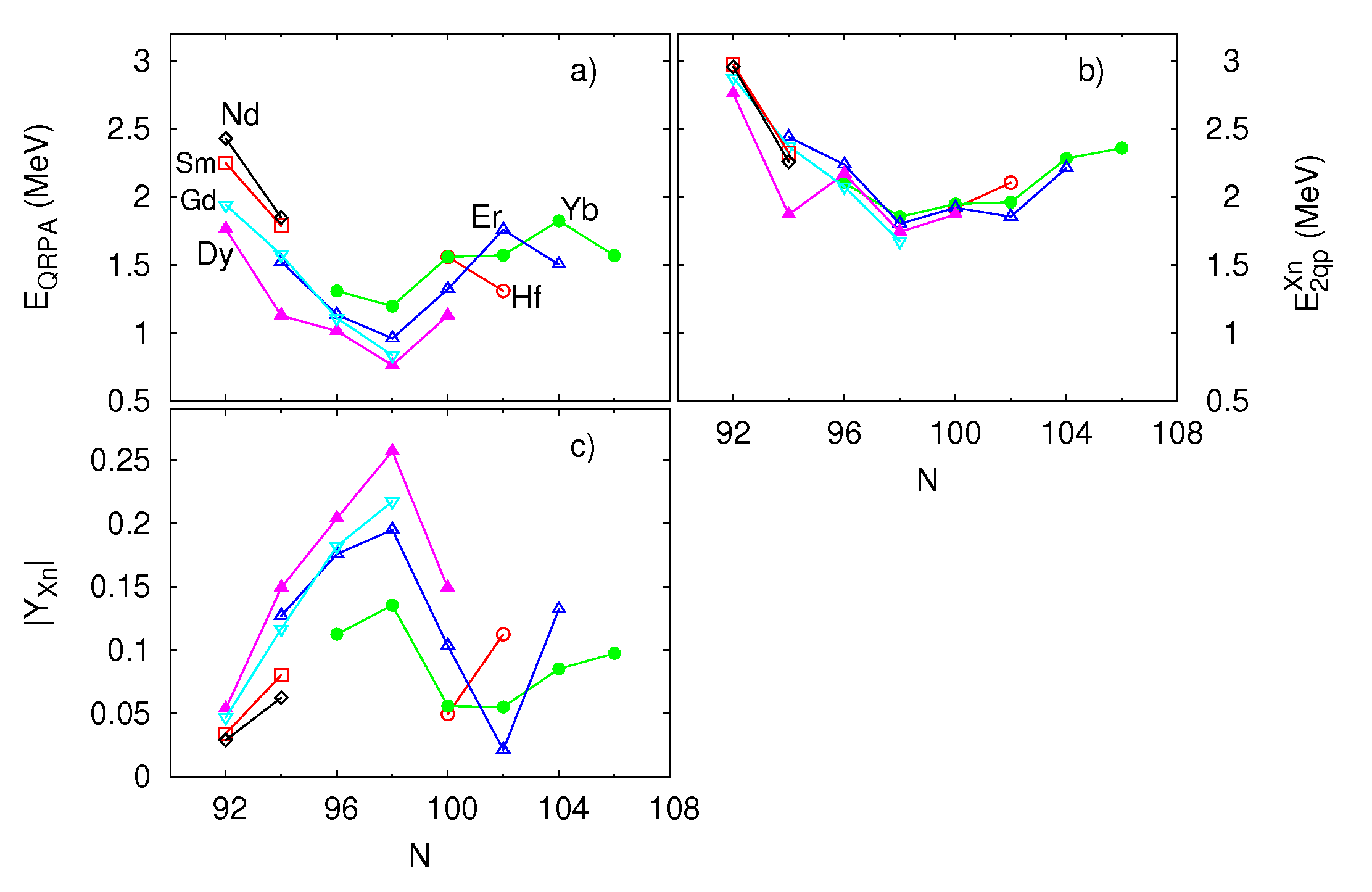}%
\caption{\label{fig_eqrpa_e2qp_y_n_2+20fm60MeVm4m8} (Color online) For the
functional SkM$^\ast$, a) Calculated $\gamma$-vibrational energy, b)
two-quasiparticle energy $E_\textrm{2qp}^\textrm{Xn}$ of the component with the
largest neutron forward amplitude, and c) absolute value $|Y_\textrm{Xn}|$ of
the backward amplitude of the same component, all as functions of neutron
number $N$.}
\end{figure*}

\begin{figure*}
\includegraphics[width=12cm]{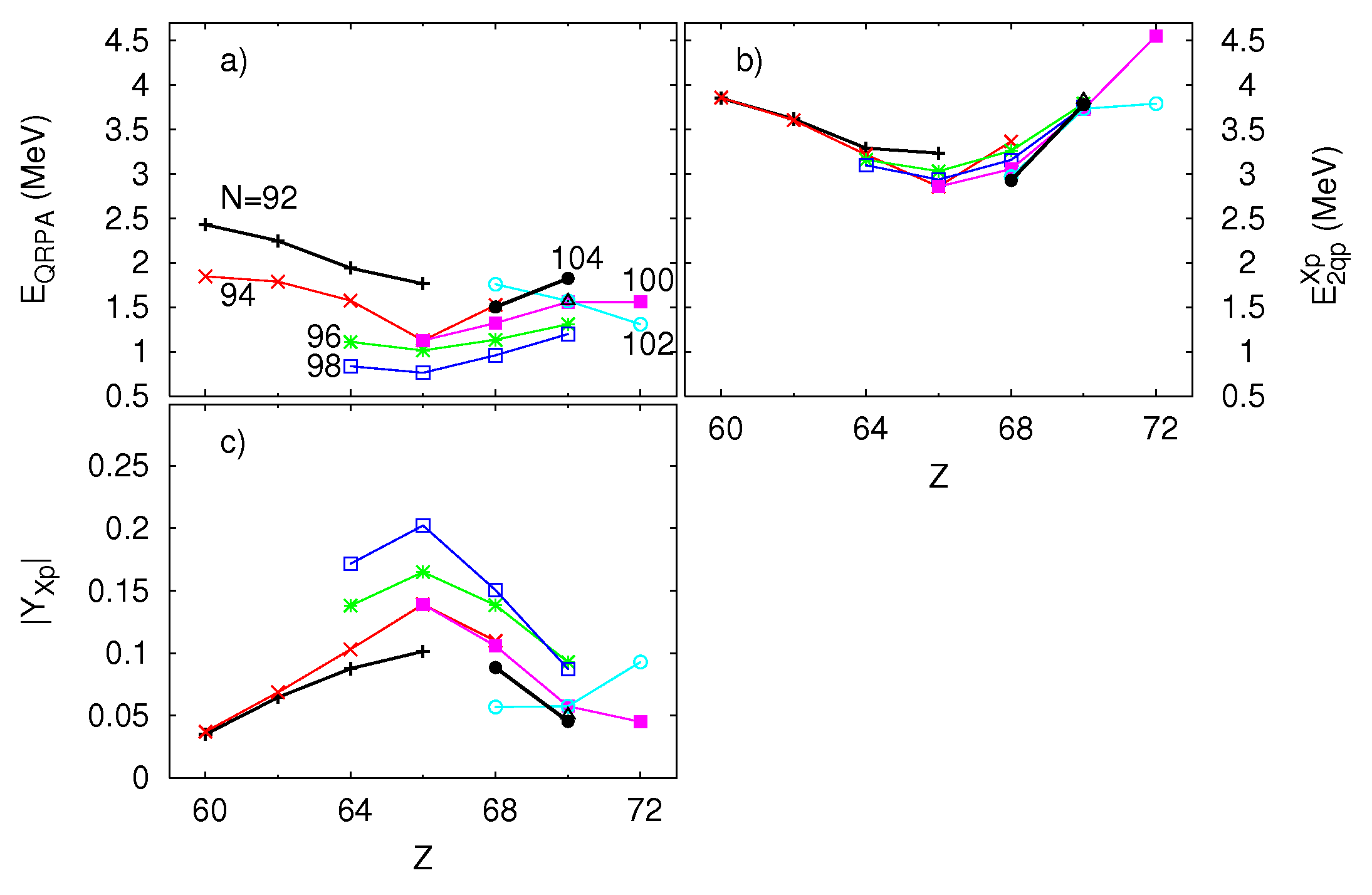}%
\caption{\label{fig_eqrpa_e2qp_y_z_2+20fm60MeVm4m8} (Color online) The same
panels as Fig.~\ref{fig_eqrpa_e2qp_y_n_2+20fm60MeVm4m8} but for protons rather
than neutrons. The connected points are isotones, and the same symbol indicates
a given isotone in each panel. $E_\textrm{2qp}^\textrm{Xp}$ and
$|Y_\textrm{Xp}|$ now refer to the proton components.}
\end{figure*}

Our calculations show a stronger dependence on $N$ than do the data in most
isotopic chains, behavior that may be due to insufficient configuration mixing
in our calculations.  Figure \ref{fig_eqrpa_e2qp_y_n_2+20fm60MeVm4m8} shows the
$N$-dependence of the calculated $\gamma$-vibrational energy, the
two-quasiparticle energy $E_\textrm{2qp}^\textrm{Xn}$ of the component, in the
quasiparticle basis, with the largest neutron forward amplitude, and the
absolute value of the backward amplitude $|Y_\textrm{Xn}|$ of the same
component, all for SkM$^\ast$.  (We transformed amplitudes from the canonical-quasiparticle 
basis to do this analysis.) For $N \leq$ 100, The $\gamma$-vibrational energy
is positively correlated with $E_\textrm{2qp}^\textrm{Xn}$, and anticorrelated
with $|Y_\textrm{Xn}|$, indicating a connection between the $N$-dependence of
those solutions and a particular two-quasiparticle state.  The downward shift
of about 1 MeV between the two-quasiparticle energy and the full QRPA energy,
seen in panels a) and b), then characterizes the effect of the residual
interaction.  Fig.~\ref{fig_eqrpa_e2qp_y_z_2+20fm60MeVm4m8} shows all the same
phenomena in the $Z$ dependence of our results, except in the energies of the
$N=102$ isotones.
 
\begin{figure}
\includegraphics[width=7cm]{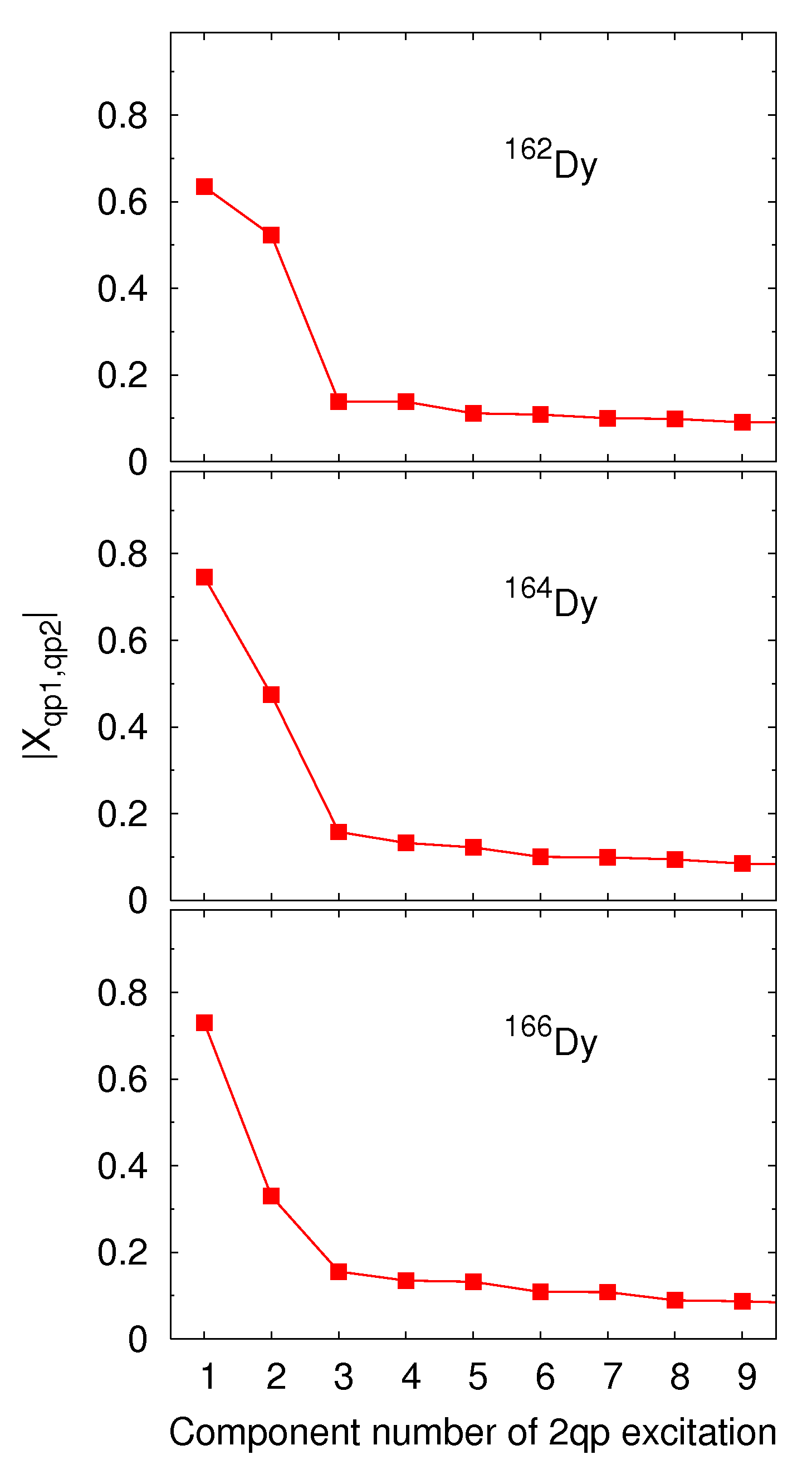}%
\caption{\label{fig_x_dy_2+20fm60MeVm4m8} (Color online) Absolute values of the
nine largest neutron forward amplitudes in $^{162-166}$Dy.  The integer on the
$x$-axis labels the two-quasiparticle components.}
\end{figure}

Figure \ref{fig_x_dy_2+20fm60MeVm4m8} shows the absolute values of the nine
largest neutron forward amplitudes in three Dy isotopes around $^{164}$Dy,
which is the one with the lowest phonon energy.  Clearly the two largest
components are far more important than the rest.  And though we don't show it,
a similar curve characterizes the protons.  From all this, we conclude that the
two-quasiparticle state with the largest neutron forward amplitudes plays a
significant role in the $N$-dependences of the QRPA solutions, and that the
same statement is true of proton forward amplitudes and $Z$ dependence.  (The
second largest components are potentially also important.) The weaker $N$- and
$Z$-dependence in the data suggests that we exaggerate the importance of those
particular two-quasiparticle states, perhaps by underestimating configuration
mixing.  It is quite possible that a better solution requires many-body
correlations beyond the QRPA. 
 
\begin{figure}
\includegraphics[width=7cm]{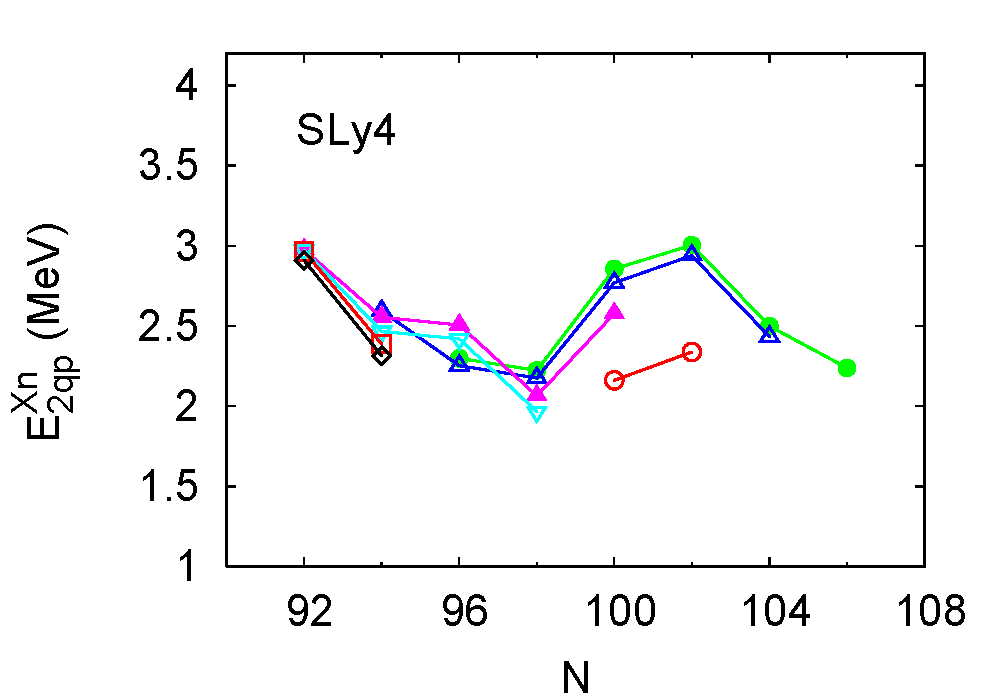}%
\caption{\label{fig_e2qp_sly4} (Color online) $E^\textrm{Xn}_\textrm{2qp}$
produced by SLy4. The symbols correspond to those of
Fig.~\ref{fig_eqrpa_e2qp_y_n_2+20fm60MeVm4m8}.}
\end{figure}

Figure \ref{fig_e2qp_sly4} shows $E^\textrm{Xn}_\textrm{2qp}$ for the SLy4
calculation. Interestingly, the range of the $E^\textrm{Xn}_\textrm{2qp}$ is
close to that produced by SkM$^\ast$, as one can see by comparing with panel b)
of Fig.~\ref{fig_eqrpa_e2qp_y_n_2+20fm60MeVm4m8}.  We conclude that the effects
of the residual interaction on $A$ dependence are quite different in the two
calculations, leading to the noticeable differences in
Fig.~\ref{fig_e_exp_skmstar_sly4_delaroche}.
 
\begin{figure*}
\includegraphics[width=12cm]{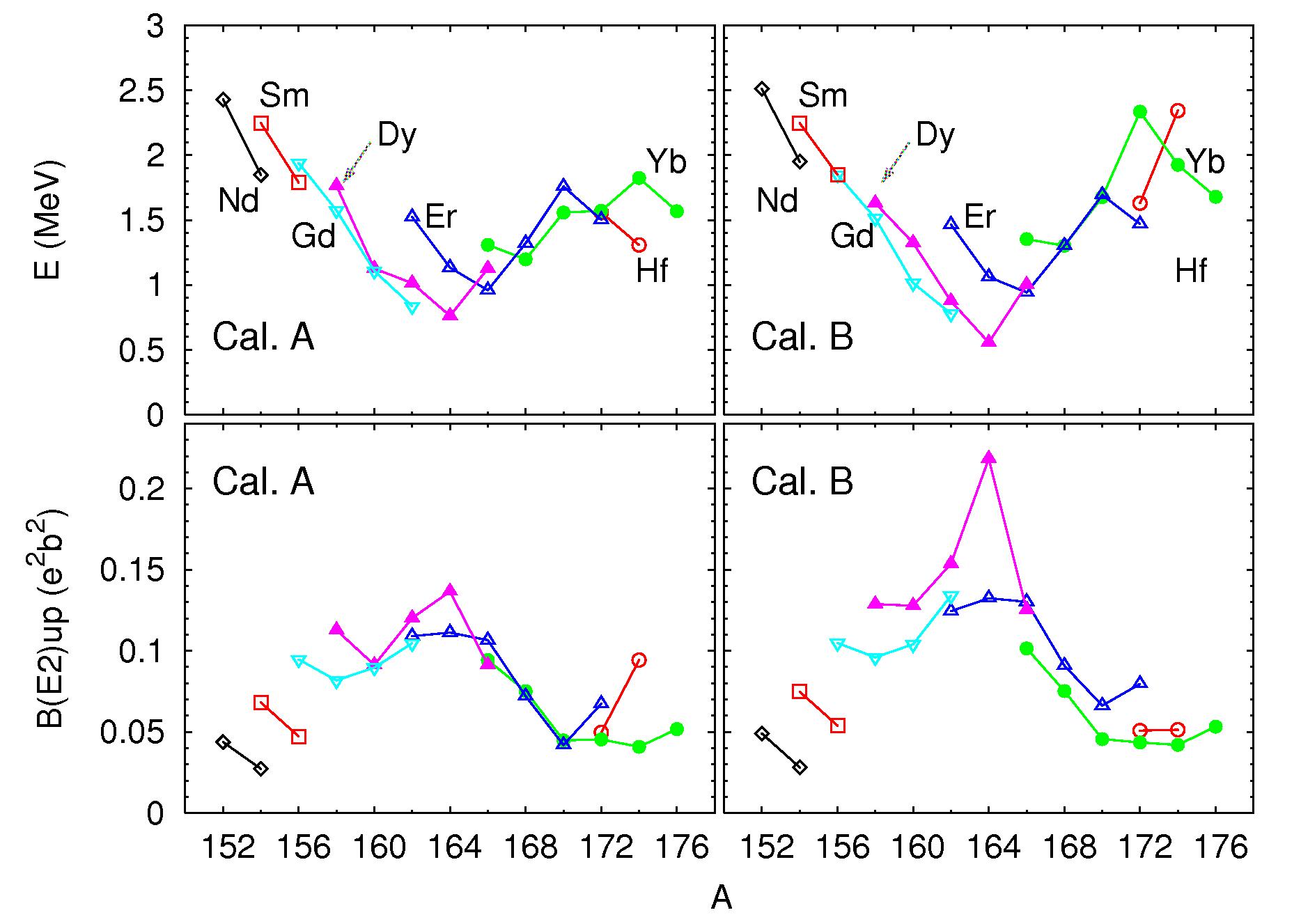}%
\caption{\label{fig_e-be2_2+20fm60MeVm4m8_2+20fm200MeVm3m6} (Color online)
Results of two calculations with different values for $E_\textrm{cut}$ and the
pairing strength $G_q$, where $q$=p (proton) or n (neutron). Calculation A
(left) uses ($E_\textrm{cut}$, $G_\textrm{n}$, $G_\textrm{p}$) = (60 MeV,
218.521 \textrm{MeV fm}$^3$, 176.364 \textrm{MeV fm}$^3$), and calculation B
(right) uses (200 MeV, 168.384 \textrm{MeV fm}$^3$, 143.996 \textrm{MeV
fm}$^3$).  The SkM$^\ast$-based results discussed in prior sections were
obtained from Calculation A.}
\end{figure*}
\begin{figure}
\includegraphics[width=7cm]{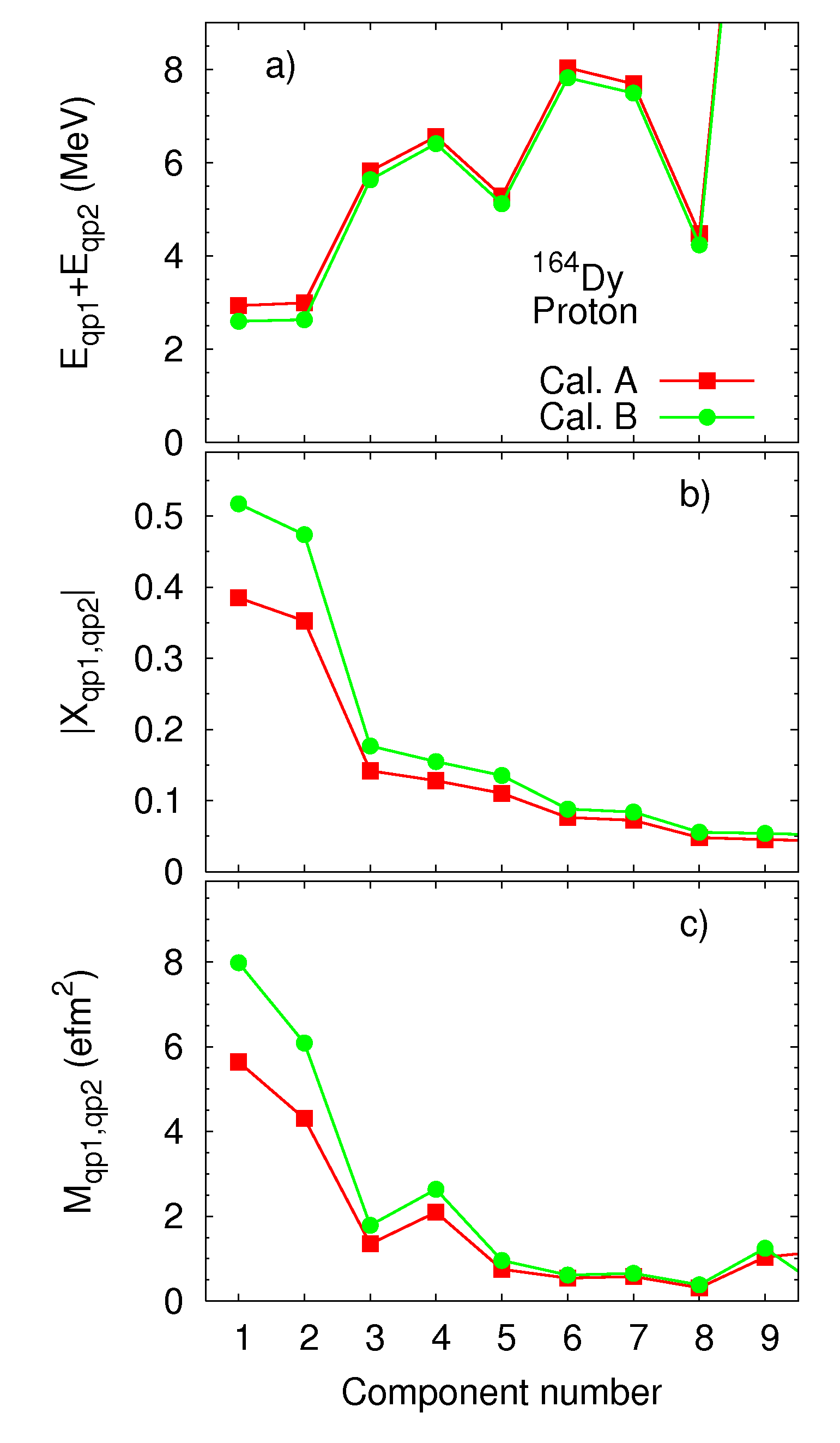}%
\caption{\label{fig_analysis_dy164} (Color online) Analysis of
two-proton-quasiparticle configurations in the $\gamma$-vibrational state of
$^{164}$Dy, with SkM$^\ast$: a) Two-quasiparticle energies, b) absolute values
of the forward amplitude, c) contribution to the transition matrix element.
The $x$ axis is the same as in Fig.~\ref{fig_x_dy_2+20fm60MeVm4m8}.  The
pairing parameters characterizing calculations A and B are given in the caption
of Fig.~\ref{fig_e-be2_2+20fm60MeVm4m8_2+20fm200MeVm3m6}.}
\end{figure}
  
\subsection{$\bm{\delta}$-pairing functional}

Low-energy quasiparticle states are obviously affected by the choice of pairing
functional, and the volume pairing we use can be varied without worsening its
ability to reproduce pairing gaps.  The reason is that the $\delta$-function
interaction is singular in the pairing channel and so must be regularized (see,
e.~g.~Ref.~\cite{Bul02}).  Here we do so by cutting off the
single-quasiparticle spectrum.  This procedure makes the strength of the
interaction depend on the cutoff as well as on the experimental pairing gaps to which it is fit. 
To illustrate the effect of the cutoff on $\gamma$
vibrations, we show in Fig.~\ref{fig_e-be2_2+20fm60MeVm4m8_2+20fm200MeVm3m6}
the results of calculations with the two different cutoffs $E_\textrm{cut}$
mentioned in Sec.\ \ref{sec:test}.  The two cutoffs require different pairing
strengths $G_q$ (for the values, see the caption) to ensure similar predictions
for pairing gaps.  We refer to the two calculations as A and B, with A having
the smaller $E_\textrm{cut}$, and therefore the larger pairing strength.

The differences in the results are mostly minor, but the $B(E2)\!\!\uparrow$ in
$^{164}$Dy, the energy and $B(E2)\!\!\uparrow$ in $^{174}$Hf, and the energy
in $^{172}$Yb are all quite different.  We account for the differences in
$^{164}$Dy by referring to Fig.~\ref{fig_analysis_dy164}.  Since calculation A
uses a larger pairing strength, it produces higher energies for low-lying
quasiparticles than does calculation B.  In the separable approximation
\cite{Nes06, Sev08}, the forward QRPA amplitudes can be written as
\begin{equation}
X_\textrm{ qp1,qp2} \propto 1/(E_\textrm{qp1}+E_\textrm{qp2} - E_\gamma) \,,
\label{eq_x_separable_approximation} 
\end{equation}
where qp1 and qp2 denote quasiparticle states, and $E_\gamma$ is the energy of
the $\gamma$-vibrational state. Using Eq.~(\ref{eq_x_separable_approximation})
and the values read from the figures, one can estimate the ratio of the forward
amplitudes of the largest two-quasiprarticle component in calculations A and B.
The result, under the assumption that the interaction matrix
elements are the same in the two calculations, is
\begin{equation}
\bigg|\frac{X_\textrm{qp1,qp2}^A}{X_\textrm{qp1,qp2}^B}\bigg| 
\simeq 0.9 \,.
\end{equation}
Panel b of Fig.~\ref{fig_analysis_dy164} shows that the exact ratio is 0.8, so
that half the difference between the two calculations can be  explained by considering only the quasiparticle energies.  This analysis implies that the QRPA solution
is sensitive to the energies of important quasiparticle states, and thus to the
pairing functional, when those energies are small.  One can take advantage of
this to fix the cutoff energy as well as the pairing strength by fitting to
properties that depend sensitvely on low-energy quasiparticle states.  Our
calculation shows, for example, that $E_\textrm{cut}=60$ MeV is better than 200
MeV.

The differences between calculations A and B in $^{172}$Yb and $^{174}$Hf are
more complicated (the corresponding two-quasiparticle components do not have
the same order as Fig.~\ref{fig_analysis_dy164}), and we could not find a
simple explanation for them.  And changes in the pairing cutoff are clearly not
enough to fix the problem with $N$- and $Z$-dependence in the QRPA.  

\subsection{Comparison with older calculations of $\bm{\gamma}$-vibrational states}

Early work on vibrations in rare-earth nuclei often made use of the
pairing-plus-QQ (quadrupole-quadrupole) Hamiltonian, both in the (Q)RPA
\cite{Mar63,Bes65,Zie71,Ham83,Mat87} and in approximations that went beyond the
QRPA order, e.~g.~\cite{Kis76,Sol89,Sol93}.  Single-particle energies were
usually obtained from the Nilsson potential, with slight shifts to improve
phenomenology, and the strength of the QQ interactions was modified slightly
from the self-consistent value so as to reproduce the energies of the
$\gamma$-vibrational states.  The adjustment to energies means that
$B(E2)\!\!\uparrow$'s are an important test of the model's predictive power.
\begin{figure}
\includegraphics[width=7cm]{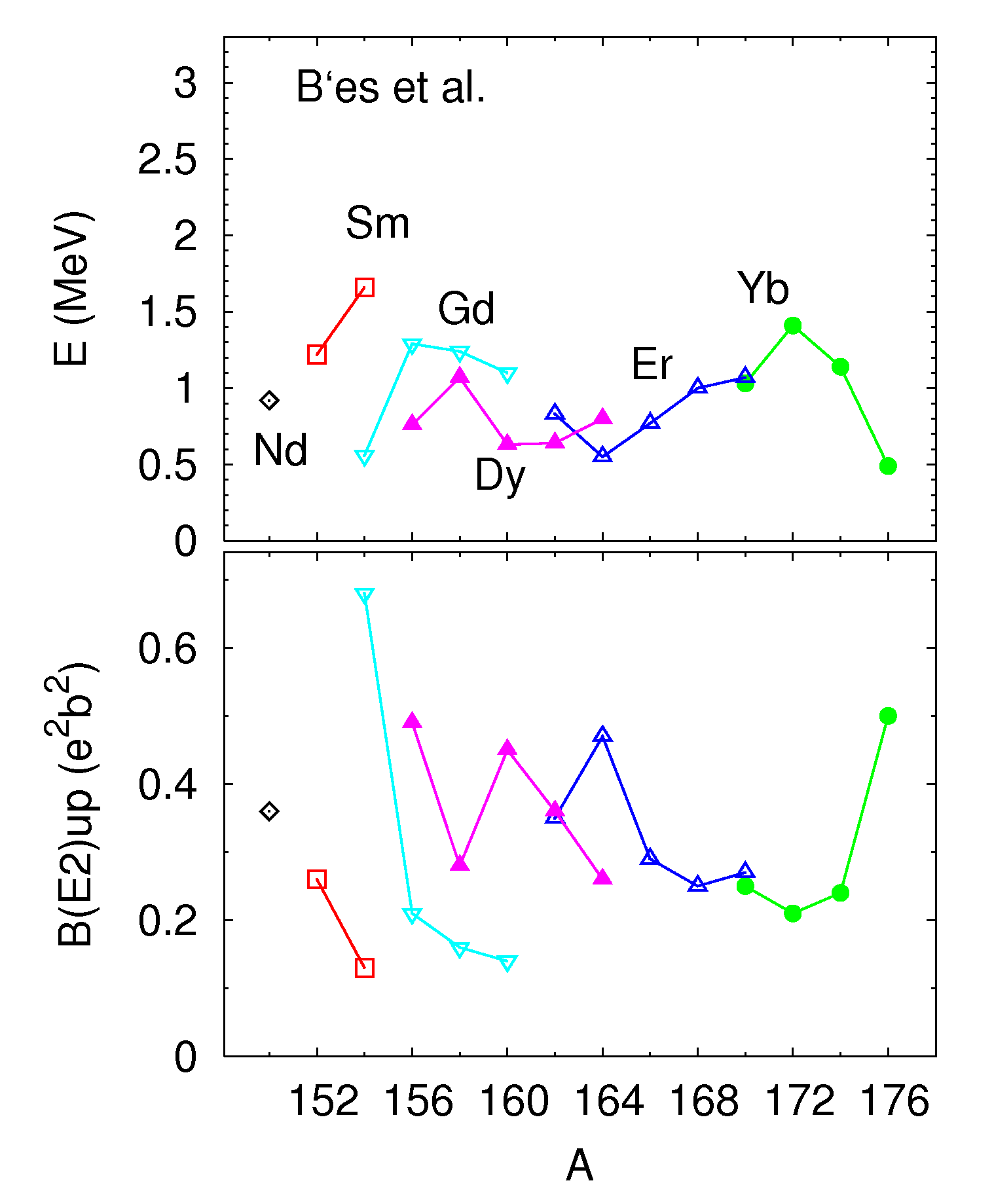}%
\caption{\label{fig_e-be2_Bes} (Color online) Energy (upper panel), and
$B(E2)\!\!\uparrow$ (lower panel) of $\gamma$-vibrational states in
calculations by D.~R.~B{\`{e}}s et al.~\cite{Bes65}} 
\end{figure}

Figure \ref{fig_e-be2_Bes} shows the energies and $B(E2)\!\!\uparrow$ from
Ref.~\cite{Bes65}.  Their energies were perhaps not quite as good overall as
ours, but also did not exhibit the sharp minimum we get around $A\sim$ 164.
The authors themselves stated that no single interaction strength reproduces
the energies of all nuclei calculated.  Their $B(E2)\!\!\uparrow$'s are too
large by a factor of two or more, a deficiency that was pointed out again in
Refs.~\cite{Mar63,Ham83}. Marshalek et al.~\cite{Mar63} listed approximations
that might cause problems in predicted $B(E2)\!\!\uparrow$'s.  Since we do
better in that observable, we believe that the cause is in fact the
interaction.  

Rare-earth $\gamma$ vibrations have also been addressed in other models.
Reference \cite{Kis76}, using the boson-expansion method, obtained
$B(E2)\!\!\uparrow$ = 0.130 $e^2$b$^2$ in $^{154}$Sm, a value somewhat larger
than ours.  Soloviev et al.~\cite{Sol89,Sol93} used the quasiparticle-phonon
nuclear model, which includes two-phonon couplings, and obtained
$B(E2)\!\!\uparrow$ = 0.127 $e^2$b$^2$ ($^{168}$Er), 0.042 $e^2$b$^2$
($^{172}$Yb), and 0.122 $e^2$b$^2$ ($^{178}$Hf). Those transition probabilities
are close to the experimental data (0.116 $e^2$b$^2 $in $^{178}$Hf
\cite{Sol89}), and the fit of the interaction meant that energies were also
reproduced well.  See also Ref.~\cite{Mat87} which used a modified QQ and an
effective three-body interactions.

Explicitly collective models have also been used.  Kumar \cite{Kum67} obtained
an energy of 1.438 MeV (close to the measured value) and a $B(E2)\!\!\uparrow$
of 0.163 $e^2$b$^2$ for the $\gamma$-vibrational state of $^{154}$Sm by solving
the Schr{\"{o}}dinger equation in collective quadrupole degrees of freedom
(Bohr and Mottelson's collective model) with the Myers-Swiatechi potential.
Garc{\'{i}}a-Ramos et al.~\cite{Gar03} used the interacting boson model (IBM)
to obtain low-energy states in about 20 even-even rare-earth nuclei, eight of
which are discussed here.  For each isotopic chain they determined the
parameters of the IBM Hamiltonian by approximately reproducing the measured
excitation energies for $J^\pi$ = 2$^+_1$, 4$^+_1$, 6$^+_1$, 8$^+_1$, 0$^+_2$,
2$^+_3$, 4$^+_3$, 2$^+_2$, 3$^+_1$, and 4$^+_2$ states, and determined the
boson effective charge by reproducing several measured $B(E2)$'s.  With regard
to the $\gamma$-vibrational $B(E2)\!\!\uparrow$'s of the eight nuclei computed
here, they reproduced those of $^{158,160}$Gd well but overestimated others.
See also Ref.~\cite{War83}, which presented another set of IBM calculations.
 
\section{\label{sec:beta} $\bm{\beta}$-vibrations}

In Tab.~\ref{tab_beta-vib_4nuclei} we show calculated and measured energies and
$B(E2)\!\!\uparrow$'s, with SkM$^\ast$, for ``$\beta$-vibrational'' states.  As
mentioned earlier, the $K^\pi=0^+$ channel contains a spurious state, and we
display only those ``$\beta$-vibrations'' that are clearly uncontaminated by
spurious motion (see the last column of the table).
\begin{table}[h]
\caption{\label{tab_beta-vib_4nuclei}%
Properties of ``$\beta$-vibrational'' states in four nuclei.
$E_\textrm{cal}^\beta$ and $E_\textrm{exp}^\beta$ are the calculated (with SkM$^\ast$) 
and experimental energies. $B(E2)\!\!\uparrow_\textrm{cal}^\beta$ and
$B(E2)\!\!\uparrow_\textrm{exp}^\beta$ are the corresponding reduced upward
transition probabilities. The ``Ratio of $S_N$'' is the ratio of the
(spurious) strength associated with the particle-number operator \cite{Ter10}
for the ``$\beta$-vibrational'' state to that for the spurious state (average
of proton and neutron). }
\begin{ruledtabular}
\begin{tabular}{cccccc}
 Nucleus & $E_\textrm{cal}^\beta$ & $E_\textrm{exp}^\beta$ &
 $B(E2)\!\!\uparrow_\textrm{cal}^\beta$ &
 $B(E2)\!\!\uparrow_\textrm{exp}^\beta$ & Ratio of $S_N$ \\
	 & (MeV) & (MeV) & $(e^2\textrm{b}^2)$ & $(e^2\textrm{b}^2)$ & \\
\colrule $^{166}$Yb & 1.802 & 1.043 & 0.0398 & & 0.004 \\
 $^{168}$Yb & 2.039 & 1.155 & 0.0343 &            & 0.012 \\
 $^{172}$Yb & 1.605 & 1.117 & 0.0049 & 0.0081(17) & 0.054 \\
 $^{170}$Er & 1.596 & 0.960 & 0.0030 & 0.0079(9)\mbox{\ \ $\!$} & 0.054 \\
\end{tabular}
\end{ruledtabular}
\end{table}

We obtained these results with calculation B (see above), the large cutoff in
which should lead to a more accurate treatment of the spurious state, though
contamination in nuclei not shown in the table indicates that a finer mesh is
necessary with a large cutoff.  In our previous paper \cite{Ter10}, which used
$E_\textrm{cut}$ = 60 MeV, $E_\textrm{cal}^\beta$ and
$B(E2)\!\!\uparrow_\textrm{cal}^\beta$ were 1.390 MeV and 1.117
$e^2\textrm{b}^2$ in $^{172}$Yb; our new energy is thus 15\% larger.  Compared
to the $\gamma$-vibrational states, overall we apparently overestimate energies
and underestimate $B(E2)\!\!\uparrow$'s, and do not do as good a job as with
$\gamma$-vibrational sates. Reference \cite{Gar01} points out that
``$\beta$-vibrational'' states are not purely vibrational, and in many cases
are better interpreted as the second member of the $K^\pi$=0$^+$ yrare
rotational band.  The QRPA cannot describe rotational bands and so the
discrepancy between our results and experiment is not totally surprising.
 
\section{\label{sec:conclusion} Conclusion}

We have used the QRPA with the Skyrme functionals SkM$^\ast$ and SLy4 and
volume-$\delta$-pairing to calculate the energies and $B(E2)\!\!\uparrow$'s of
$\gamma$-vibrational states in well-deformed even-even rare-earth nuclei.
SkM$^\ast$ proves to be the better functional.  The range of calculated values
overlaps well with that of the experimental data. Since the QRPA energies are
appreciably different from their unperturbed counterparts, that counts as a
success for the residual interaction; the vibrational states discussed here are
not taken into account at all in determining energy functionals.  In detail,
however, the calculations are far from perfect, and their $N$- and $Z$-
dependence suggest the importance of many-body correlations that are not
included in the QRPA.  And our representation of ``$\beta$-vibrational'' states
turns to be worse than that of $\gamma$ vibrations, probably because ``$\beta$
vibrations'' are often not really vibrations.


We also suggested that the cutoff associated with volume pairing can be fixed
along with the pairing strength by examining properties that are sensitive to the
structure of low-lying quasiparticles.  

Our calculation is better overall than the works of half a century ago.  The
aims of the pairing-plus-QQ model are much more limited than those of nuclear
DFT; the mean field arising from pairing-plus-QQ is an infinitely deep well,
and so the model cannot make predictions for binding energies or for excitation
energies near the drip line (where the underlying Nilsson single-particle
potential is not appropriate).  Despite the increasing sophistication of the
many-body methods that have been applied together with the pairng-plus-QQ
model, a more general framework such as DFT appears necessary for the unified
description of heavy nuclei.

Finally, we have shown that in this era of supercomputing a scalable code makes
systematic and fully self-consistent Skyrme-QRPA studies possible.  We expect,
as a result, that excited states will play an increasing role in the
determination of nuclear density functionals.
 
\begin{acknowledgments}
We thank Drs.~Umar and Oberacker for letting us to use their HFB code.  This
work was supported by the UNEDF SciDAC Collaboration under DOE Grant No.\
DE-FC02-07ER41457 and by the National Science Foundation through Teragrid
resources provided by the National Institute for Computational Sciences. We
also used computers at the National Energy Research Scientific Computing Center.
\end{acknowledgments}
 
\appendix*

\section{Coulomb-direct matrix elements}

The computation of the direct two-body matrix elements of the Coulomb
interaction consumes a lot of computing time.  In this appendix we present our
implementation of that computation.

The Coulomb interaction is 
\begin{equation} 
V_\textrm{C}(\bm{r}_1,\bm{r}_2) =
\frac{e^2}{\lvert \bm{r}_1 - \bm{r}_2 \rvert } \,. \label{eq_coul_int}
\end{equation} We take advantage of axial symmetry to write the wave function
as \begin{equation} X_a(\bm{r}) = \frac{1}{ \sqrt{2\pi} }\sum_{\sigma=\pm{1/2}}
{\cal F}_a(\sigma; z,\rho)e^{i(j^z_a-\sigma)\phi}\vert\sigma \rangle~,
\label{eq_wave_function} 
\end{equation} 
The label $a$ stands for $(q\pi j_z i)$, i.~e.\ particle type, parity,
angular-momentum $z$-component, and an additional label $i$ to fully specify
the state.  The position $\bm{r}$ is represented in cylindrical coordinates,
and the label $\sigma=\pm 1/2$ is the $z$-component of the spin. The function
${\cal F}_a(\sigma;z,\rho)$ is treated numerically.  The set $\{X_a(\bm{r})\}$
can refer to any single-particle basis (or components of quasiparticle basis
states, in which case another label to distinguish upper from lower is
necessary) with axial and parity symmetries.  In our calculations we use the
canonical single-particle basis. 

With the help of a few well-known formulae from Appendix B of Ref.\
\cite{Mes61}, one can obtain the expansion
\begin{eqnarray}
\frac{1}{\vert \bm{r}_1 - \bm{r}_2 \vert} 
 &&
=\sum_{l=0}^\infty \frac{ (\sqrt{\rho_<^2+z_<^2})^l }
{ (\sqrt{\rho_>^2+z_>^2})^{l+1} }
\sum_{m=-l}^l\frac{ (l-m)! }{ (l+m)! } \nonumber \\
&&
P_{lm}\bigglb( \frac{ z_1 }{ \sqrt{\rho_1^2+z_1^2} } \biggrb)
P_{lm}\bigglb( \frac{ z_2 }{ \sqrt{\rho_2^2+z_2^2} } \biggrb)
\nonumber \\
&& 
e^{im(\phi_2-\phi_1)}~, \label{eq_expansion_1ovr1r2}
\end{eqnarray}
where
\begin{eqnarray}
\begin{array}{l}
\rho_<^2+z_<^2 = \rho_1^2+z_1^2 \\
\rho_>^2+z_>^2 = \rho_2^2+z_2^2
\end{array}
\Bigr\}~,
~\textrm{if}~\rho_2^2+z_2^2 > \rho_1^2+z_1^2~, \nonumber \\
\begin{array}{l}
\rho_<^2+z_<^2 = \rho_2^2+z_2^2 \\
\rho_>^2+z_>^2 = \rho_1^2+z_1^2
\end{array}
\Bigr\}~,
~\textrm{if}~\rho_2^2+z_2^2 < \rho_1^2+z_1^2~, \label{eq_def_rholarge_rhosmall}
\end{eqnarray}
and the $P_{lm}$ are associated Legendre polynomials \cite{Mes61}.  By using
Eqs.~(\ref{eq_coul_int})$-$(\ref{eq_def_rholarge_rhosmall}), one can then write
the matrix element of the Coulomb-direct interaction as
\begin{widetext}
\begin{eqnarray}
V^\textrm{C}_{ab,cd} 
&=& \int d^3\bm{r}_1 \int d^3\bm{r}_2
X_a^\dagger(\bm{r}_1) X_b^\dagger(\bm{r}_2)
V_\textrm{C}(\bm{r}_1,\bm{r}_2) X_c(\bm{r}_1) X_d(\bm{r}_2) \nonumber \\
&=& e^2 \sum_{\sigma_a,\sigma_b} \sum_{l=0}^\infty
\sum_{m=-l}^l \delta_{-j^z_a-m+j^z_c,0}\delta_{-j^z_b+m+j^z_d,0}
\int_{-\infty}^\infty dz_1 \int_0^\infty d\rho_1 \rho_1
\int_{-\infty}^\infty dz_2 \int_0^\infty d\rho_2 \rho_2
{\cal F}_a(\sigma_a;z_1,\rho_1) \nonumber \\
&&P_{lm}\bigglb( \frac{z_1}{\sqrt{\rho_1^2+z_1^2}} \biggrb)
{\cal F}_c(\sigma_a;z_1,\rho_1)
{\cal F}_b(\sigma_b;z_2,\rho_2)
P_{lm}\bigglb( \frac{z_2}{\sqrt{\rho_2^2+z_2^2}} \biggrb)
{\cal F}_d(\sigma_b;z_2,\rho_2)
\frac{ \biglb( \sqrt{\rho_<^2+z_<^2} \bigrb)^l }{ \biglb( \sqrt{\rho_>^2+z_>^2} \bigrb)^{l+1} } \nonumber \\
&&\frac{(l-m)!}{(l+m)!}~. \label{eq_VC_1}
\end{eqnarray}
Changing variables to 
\begin{equation}
(z,R=\sqrt{\rho^2+z^2})~,
\end{equation}
and noting that 
\begin{eqnarray}
&&{\cal F}_a(\sigma_a;-z,\rho)=(-)^{j^z_a-\sigma_a}\pi_a
{\cal F}_a(\sigma_a;z,\rho)~, \\
&&P_{lm}(-x) = (-)^{l-m}P_{lm}(x)~,
\end{eqnarray}
we arrive at 
\begin{eqnarray}
V^\textrm{C}_{ab,cd} 
&=&4e^2\delta_{-j^z_a+j^z_c,j^z_b-j^z_d}
\delta_{\pi_a\pi_c,\pi_b\pi_d}
\sum_{\sigma_a,\sigma_b}\sum_{l=\lvert -j^z_a+j^z_c\rvert}^\infty \delta_{\pi_a\pi_c,(-)^l}
\frac{(l-m)!}{(l+m)!}
\bigglb\{\int_0^\infty dR_1 \frac{ {\cal T}_1(R_1) }{ R_1^{l-1} }
\int_0^{R_1} dR_2 R_2^{l+2} {\cal T}_2(R_2) \nonumber \\
&& +\int_0^\infty dR_1 {\cal T}_1(R_1) R_1^{l+2}\int_{R_1}^\infty dR_2 \frac{ {\cal T}_2(R_2) }{ R_2^{l-1} } \biggrb\}
\biggrb\rvert_{m=-j^z_a+j^z_c}~, 
\label{eq_VC_final}
\end{eqnarray}
where
\begin{eqnarray}
&&
{\cal T}_1(R_1) = \frac{1}{R_1}\int_0^{R_1}dz_1
{\cal F}_a\Biglb(\sigma_a;z_1,\sqrt{ R_1^2-z_1^2 } 
\Bigrb)
P_{lm}\bigglb( \frac{z_1}{R_1} \biggrb)
{\cal F}_c\Biglb(\sigma_a;z_1,\sqrt{ R_1^2-z_1^2 } \Bigrb)~, \nonumber \\
&&
{\cal T}_2(R_2) = \frac{1}{R_2}\int_0^{R_2}dz_2
{\cal F}_b\Biglb(\sigma_b;z_2,\sqrt{ R_2^2-z_2^2 } 
\Bigrb)
P_{lm}\bigglb( \frac{z_2}{R_2} \biggrb)
{\cal F}_d\Biglb(\sigma_b;z_2,\sqrt{ R_2^2-z_2^2 } \Bigrb)~.
\label{eq_calT}
\end{eqnarray}
\end{widetext}
Though it is not explicit in the notation, ${\cal T}_1(R_1)$ depends on the
labels $a$, $c$, $\sigma_a$, and $(l,m)$, and ${\cal T}_2(R_2)$ on similar
quantities. 

Equations (\ref{eq_VC_final}) and (\ref{eq_calT}) are what we use, with minor
modifications for hole states, in our code.  We calculate ${\cal T}_1(R_1)$ and
${\cal T}_2(R_2)$ on a mesh and store them in arrays. Once this is finished,
the time to calculate Eq.~(\ref{eq_VC_final}) is determined mainly by the nest
structure of the two-fold integrals and the summation with respect to $l$. For
a system with quadrupole deformation $\beta\sim$ 0.3 the number of terms
necessary in the sum over $l$ (much fewer than 20 in practice, with only even
or only odd $l$ contributing) is much smaller than the number of mesh points in
the integration. 

If an equidistant mesh is used for integrals in which an upper or lower bound
is a variable, the computational effort to calculate the two-fold integrals in
Eq.~(\ref{eq_VC_final}) is nearly the same as that of single integrals.  Thus,
we calculate the wave functions on a new mesh by interpolating between B-spline
points, and then use Simpson's rule with three times more mesh points than
B-spline points to preserve accuracy (while still speeding up the
integration).  We have checked our procedure by using the two-body matrix
elements that it produces to calculate the Coulomb-direct energy of the HFB
ground state, which we then compared to the output of the HFB code.
%


\end{document}